\documentclass[11pt,twoside]{article}
\usepackage{./asp2014}

\aspSuppressVolSlug
\resetcounters

\begin{document}

\title{The East-Asian VLBI Network}
\author{Kiyoaki Wajima,$^1$ Yoshiaki~Hagiwara,$^2$ Tao~An,$^3$ Willem~A.~Baan,$^3$
Kenta~Fujisawa,$^4$ Longfei~Hao,$^5$ Wu~Jiang,$^3$ Taehyun~Jung,$^1$
Noriyuki~Kawaguchi,$^6$ Jongsoo~Kim,$^1$ Hideyuki~Kobayashi,$^6$ Se-Jin~Oh,$^1$
Duk-Gyoo~Roh,$^1$ Min~Wang$^7$ Yuanwei~Wu,$^6$ Bo~Xia,$^3$ and Ming~Zhang$^7$
\affil{$^1$Korea Astronomy and Space Science Institute, Daejeon, Korea;
\email{wajima@kasi.re.kr}}
\affil{$^2$Toyo University, Tokyo, Japan}
\affil{$^3$Shanghai Astronomical Observatory, Shanghai, China}
\affil{$^4$Yamaguchi University, Yamaguchi, Japan}
\affil{$^5$Yunnan Astronomical Observatory, Yunnan, China}
\affil{$^6$National Astronomical Observatory of Japan, Tokyo, Japan}
\affil{$^7$Xinjiang Astronomical Observatory, Xinjiang, China}}

\paperauthor
{Kiyoaki~Wajima}
{wajima@kasi.re.kr}{}{Korea Astronomy and Space Science Institute}{Radio Astronomy Division}{}{Daejeon}{34055}{Korea}
\paperauthor{Yoshiaki~Hagiwara}{yhagiwara@toyo.jp}{}{Toyo University}{Faculty of Letters}{}{Tokyo}{112-8606}{Japan}
\paperauthor{Tao~An}{antao@shao.ac.cn}{}{Shanghai Astronomical Observatory}{}{}{Shanghai}{200030}{China}
\paperauthor{Willem~A.~Baan}{willem.baan@shao.ac.cn}{}{Shanghai Astronomical Observatory}{}{}{Shanghai}{200030}{China}
\paperauthor{Kenta~Fujisawa}{kenta@yamaguchi-u.ac.jp}{}{Yamaguchi University}{Research Institute for Time Studies}{}{Yamaguchi}{753-8511}{Japan}
\paperauthor{Longfei~Hao}{haolongfei@ynao.ac.cn}{}{Yunnan Astronomical Observatory}{}{}{Yunnan}{650011}{China}
\paperauthor{Wu~Jiang}{jiangwu@shao.ac.cn}{}{Shanghai Astronomical Observatory}{}{}{Shanghai}{200030}{China}
\paperauthor{Taehyun~Jung}{thjung@kasi.re.kr}{}{Korea Astronomy and Space Science Institute}{Radio Astronomy Division}{}{Daejeon}{34055}{Korea}
\paperauthor{Noriyuki~Kawaguchi}{kawagu.nori@nao.ac.jp}{}{National Astronomical Observatory of Japan}{Mizusawa VLBI Observatory}{}{Tokyo}{181-8588}{Japan}
\paperauthor{Jongsoo~Kim}{jskim@kasi.re.kr}{}{Korea Astronomy and Space Science Institute}{Radio Astronomy Division}{}{Daejeon}{34055}{Korea}
\paperauthor{Hideyuki~Kobayashi}{hideyuki.kobayashi@nao.ac.jp}{}{National Astronomical Observatory of Japan}{Mizusawa VLBI Observatory}{}{Tokyo}{181-8588}{Japan}
\paperauthor{Se-Jin~Oh}{sjoh@kasi.re.kr}{}{Korea Astronomy and Space Science Institute}{Radio Astronomy Division}{}{Daejeon}{34055}{Korea}
\paperauthor{Duk-Gyoo~Roh}{dgroh@kasi.re.kr}{}{Korea Astronomy and Space Science Institute}{Radio Astronomy Division}{}{Daejeon}{34055}{Korea}
\paperauthor{Min~Wang}{wm@ynao.ac.cn}{}{Yunnan Astronomical Observatory}{}{}{Yunnan}{650011}{China}
\paperauthor{Yuanwei~Wu}{yuanwei.wu@nao.ac.jp}{}{National Astronomical Observatory of Japan}{Mizusawa VLBI Observatory}{}{Tokyo}{181-8588}{Japan}
\paperauthor{Bo~Xia}{bxia@shao.ac.cn}{}{Shanghai Astronomical Observatory}{}{}{Shanghai}{200030}{China}
\paperauthor{Ming~Zhang}{zhang.ming@xao.ac.cn}{}{Xinjiang Astronomical Observatory}{}{}{Xinjiang}{830011}{China}

\begin{abstract}

The East-Asian VLBI Network (EAVN) is the international VLBI facility in East
Asia and is conducted in collaboration with China, Japan, and Korea.
The EAVN consists of VLBI arrays operated in each East Asian country, containing
21 radio telescopes and three correlators.
The EAVN will be mainly operated at 6.7 (C-band), 8 (X-band), 22 (K-band), and
43~GHz (Q-band), although the EAVN has an ability to conduct observations at
1.6 -- 129~GHz.
We have conducted fringe test observations eight times to date at 8 and 22~GHz
and fringes have been successfully detected at both frequencies.
We have also conducted science commissioning observations of 6.7~GHz methanol
masers in massive star-forming regions.
The EAVN will be operational from the second half of 2017, providing
complementary results with the FAST on AGNs, massive star-forming regions,
and evolved stars with high angular resolution at cm- to mm-wavelengths.

\end{abstract}

\section{Introduction}

A new VLBI array in East Asia, the East-Asian VLBI Network (EAVN), is being
planned under the collaboration of the East Asia Core Observatory Association
(EACOA).
VLBI facilities are developing and operating in each East Asian country, the
Chinese VLBI Network \citep[CVN;][]{Li08} in China, the Korean VLBI Network
\citep[KVN;][]{Lee14} in Korea, the Japanese VLBI Network \citep[JVN;][]{Doi06}
and VLBI Exploration of Radio Astrometry \citep[VERA;][]{Kobayashi03} in Japan,
and various international collaboration programs, such as the KVN and VERA
Array (KaVA), is ongoing by using those facilities.
On the basis of this background, a small task force `the EAVN Tiger Team' was
organized in June 2013, consisting of 17 members from China, Japan, and Korea,
to handle various issues related to EAVN, such as promotion of VLBI test
observations, clarification of the problems for future regular operation of
EAVN, and so on.
The work of this task force is carried out as part of activities in `the East
Asian VLBI Consortium', which is one of working groups under the EACOA.

\section{Overview of EAVN}

Figure~\ref{fig:Figure1} shows an overall image of the EAVN, which consists
of 21 potential telescopes, 5 from China, 12 from Japan, and 4 from Korea,
and two correlator sites, Korea-Japan Correlation Center (KJCC) in KASI, and
Shanghai Astronomical Observatory.
Details of the telescopes and brief specifications of the EAVN are shown in
Tables~\ref{tbl:Table1} and \ref{tbl:Table2}, respectively.
In Table~\ref{tbl:Table1}, filled symbols indicate telescopes which have
participated in fringe test observations to be described in Section~3, open
squares show antennas participated in VLBI observations of methanol masers in
massive star-forming regions at 6.7~GHz conducted in 2010 and 2011
\citep{Fujisawa14}, and open and filled triangles correspond to antennas
constituting the KaVA, which is operational at 22 and 43~GHz
\citep{Matsumoto14,Niinuma14}.

\begin{table}[b!]
\caption{Potential EAVN Telescopes.}
\label{tbl:Table1}
\smallskip
\begin{center}
{\small
\begin{tabular}{llccccccc}
\tableline
\noalign{\smallskip}
Name & Array & Longitude      & Latitude       & Diameter & \multicolumn{4}{c}{Observation Band}  \\
     &       & [$^{\circ}$~N] & [$^{\circ}$~E] & [m]      & C       & X       & K       & Q       \\
\noalign{\smallskip}
\tableline
\noalign{\smallskip}
Kunming      & CVN  & 25.027 & 102.796 & 40 & $\circ$   & $\bullet$ &                  &             \\
Miyun        & CVN  & 40.558 & 116.976 & 50 &           & $\circ$   &                  &             \\
Nanshan      & CVN  & 40.399 & 116.239 & 26 & $\circ$   & $\bullet$ & $\bullet$        &             \\
Sheshan      & CVN  & 31.099 & 121.200 & 25 & $\square$ & $\bullet$ & $\bullet$        &             \\
Tianma       & CVN  & 31.092 & 121.136 & 65 & $\circ$   & $\bullet$ & $\circ$          & $\circ$     \\
Gifu         & JVN  & 35.468 & 136.737 & 11 &           &           & $\circ$          &             \\
Hitachi      & JVN  & 36.697 & 140.692 & 32 & $\square$ & $\bullet$ & $\bullet$        &             \\
Kashima      & JVN  & 35.956 & 140.660 & 34 &           & $\circ$   & $\circ$          & $\circ$     \\
Takahagi     & JVN  & 36.699 & 140.695 & 32 &           &           & $\bullet$        &             \\
Tomakomai    & JVN  & 42.674 & 141.597 & 11 &           &           & $\circ$          &             \\
Tsukuba      & JVN  & 36.103 & 140.089 & 32 &           & $\bullet$ & $\circ$          &             \\
Usuda        & JVN  & 36.132 & 138.363 & 64 & $\square$ & $\circ$   &                  &             \\
Yamaguchi    & JVN  & 34.216 & 131.557 & 32 & $\square$ & $\bullet$ & $\circ$          &             \\
Iriki        & VERA & 31.748 & 130.440 & 20 & $\square$ & $\bullet$ & $\blacktriangle$ & $\triangle$ \\
Ishigakijima & VERA & 24.412 & 124.171 & 20 & $\square$ & $\bullet$ & $\blacktriangle$ & $\triangle$ \\
Mizusawa     & VERA & 39.134 & 141.133 & 20 & $\square$ & $\bullet$ & $\blacktriangle$ & $\triangle$ \\
Ogasawara    & VERA & 27.092 & 142.217 & 20 & $\square$ & $\bullet$ & $\blacktriangle$ & $\triangle$ \\
Sejong       & ---  & 36.520 & 127.303 & 22 &           & $\bullet$ &                  &             \\
Tamna        & KVN  & 33.289 & 126.460 & 21 &           &           & $\blacktriangle$ & $\triangle$ \\
Ulsan        & KVN  & 35.546 & 129.250 & 21 & $\circ$   & $\bullet$ & $\blacktriangle$ & $\triangle$ \\
Yonsei       & KVN  & 37.565 & 126.941 & 21 &           &           & $\blacktriangle$ & $\triangle$ \\
\noalign{\smallskip}
\tableline\
\end{tabular}
}
\end{center}
\end{table}

\begin{table}[ht!]
\caption{Specifications of EAVN.}
\label{tbl:Table2}
\smallskip
\begin{center}
{\small
\begin{tabular}{ll}
\tableline
\noalign{\smallskip}
Number of (potential) telescopes        & 21 \\
Frequency coverage                      & 6.7~GHz (12 stations), 8~GHz (16), 22~GHz (17),     \\
                                        & 43~GHz (9)                                          \\
Angular resolution                      & 1.5~mas (8~GHz), 0.6~mas (22~GHz), 0.7~mas (43~GHz) \\
7-$\sigma$ Fringe detection sensitivity & 1.6~mJy (8~GHz), 9.5~mJy (22~GHz)                   \\
                                        & (for a continuum source, $\tau = 60$~s)             \\
Recording rate                          & $\ge 1$~Gbps ($B = 256$~MHz)                        \\
Correlator                              & Korea-Japan Joint VLBI Correlator (KASI),           \\
                                        & DiFX (KASI/SHAO)                                    \\
\noalign{\smallskip}
\tableline\
\end{tabular}
}
\end{center}
\end{table}

\articlefigure
{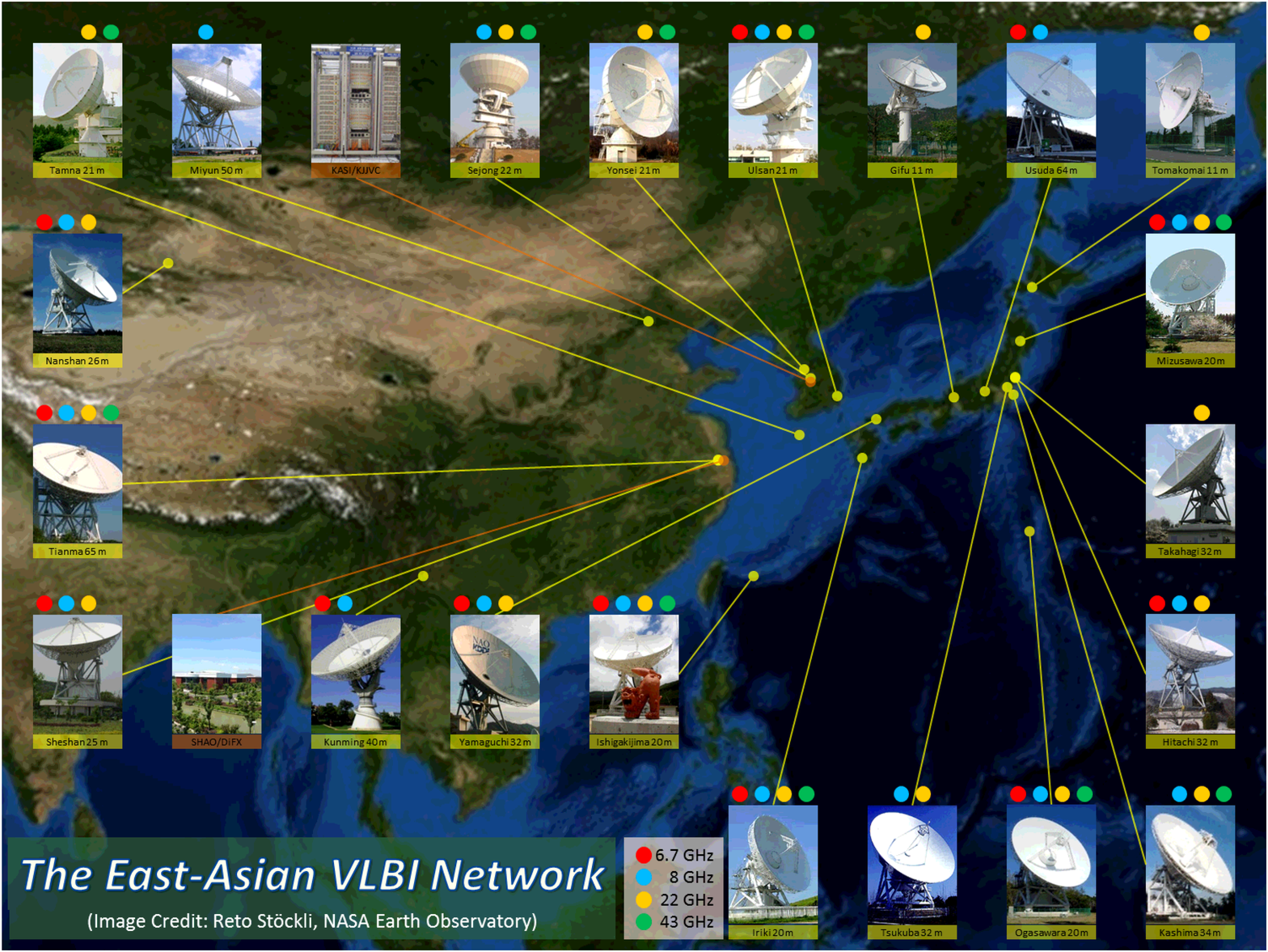}
{fig:Figure1}
{Location of EAVN sites. Photos of potential EAVN telescopes (yellow points)
and correlator sites (brown points) are overlaid on `the Blue Marble' image
(image credit: NASA's Earth Observatory).}

Figure~\ref{fig:Figure2} shows examples of ($u$, $v$) coverage for an EAVN
observation at 22~GHz with the various declination of a source.
The EAVN can sample visibilities of a wide range of baseline lengths from
20~km (JVN-Tsukuba -- JVN-Kashima) to 5,000~km (CVN-Nanshan -- VERA-Ogasawara),
allowing us to obtain a round-shaped synthesized beam with very low noise
level thanks to a lot of large antennas, as shown in Table~\ref{tbl:Table1}.

\articlefigurefour
{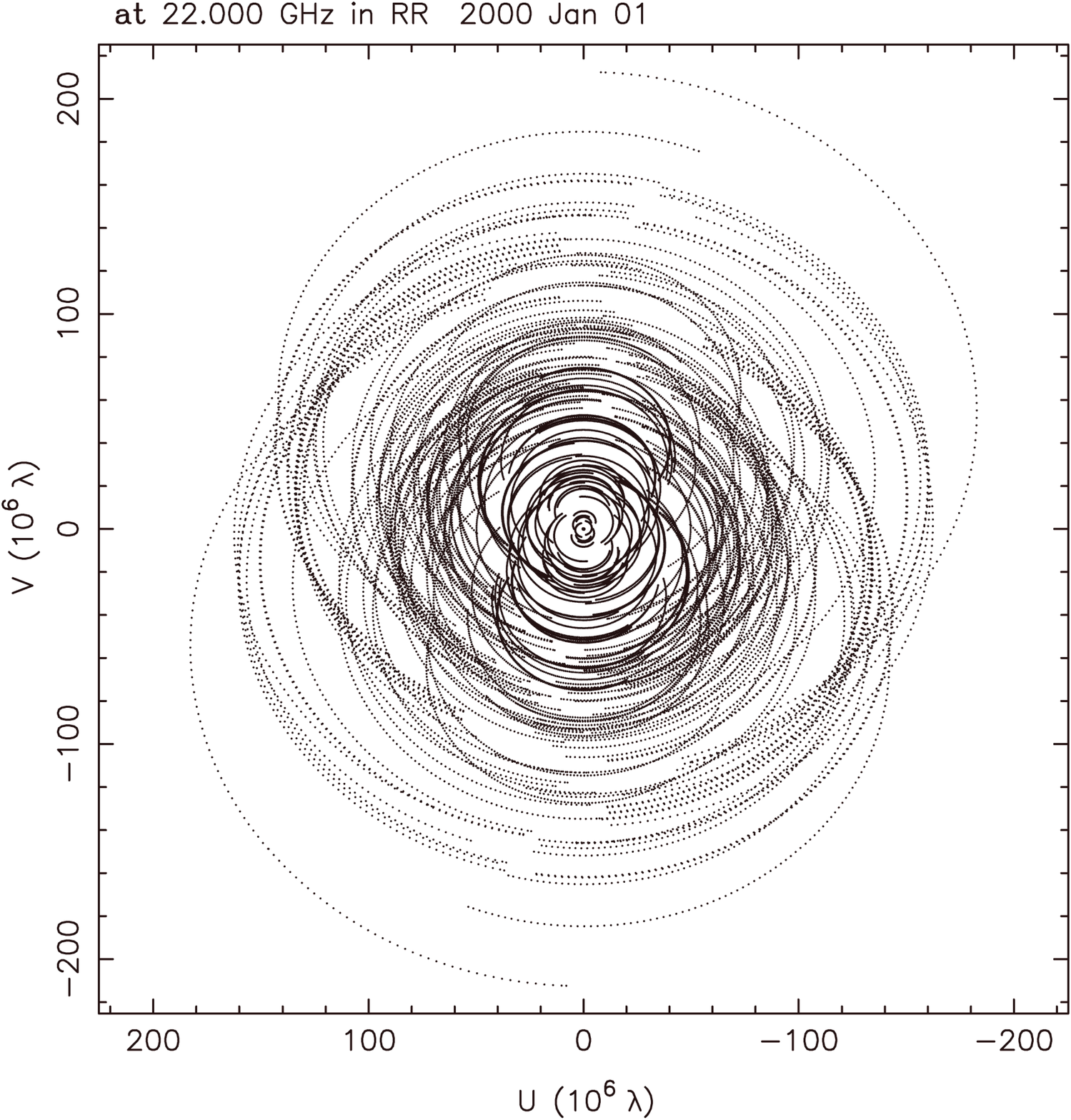}
{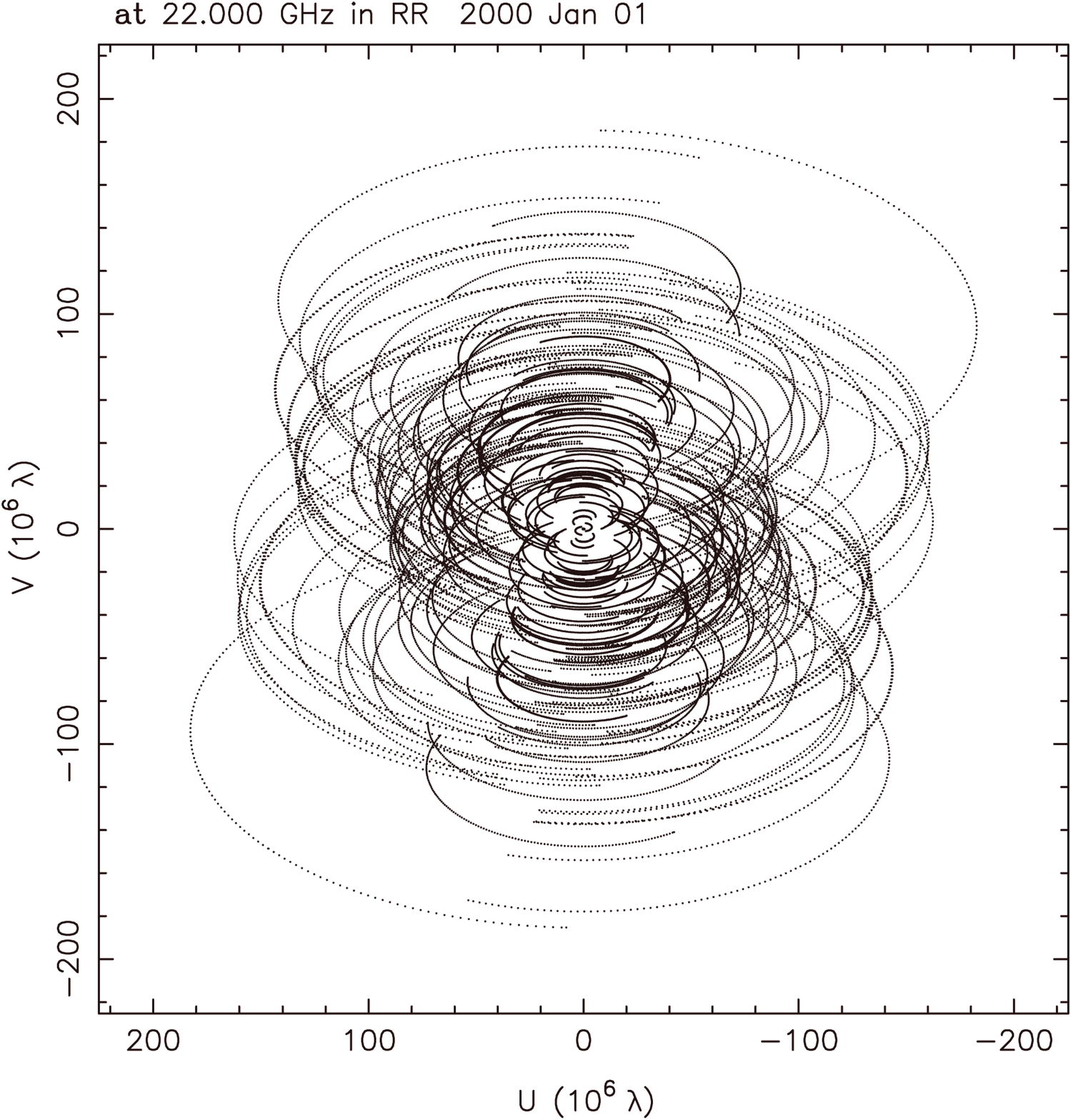}
{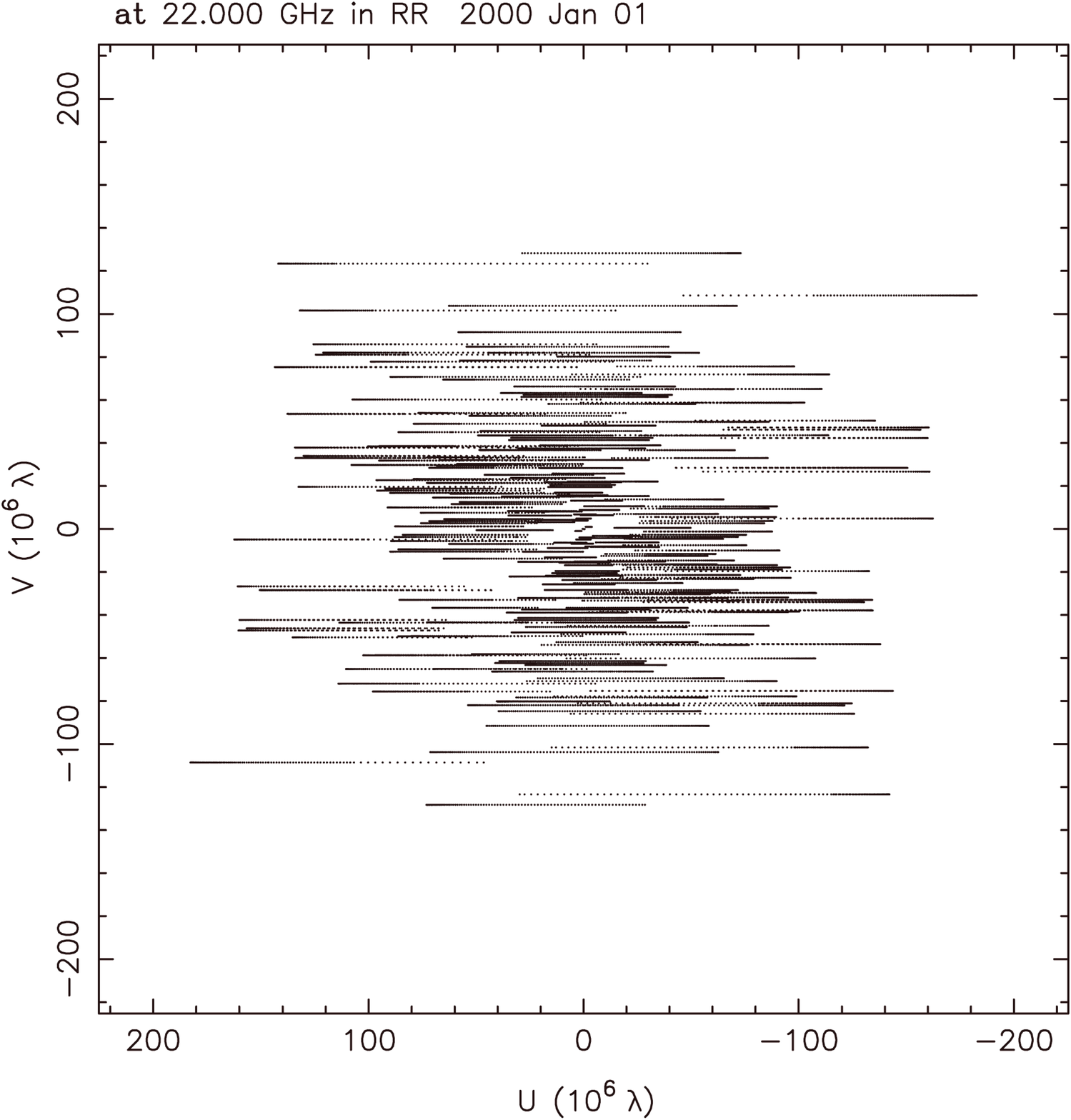}
{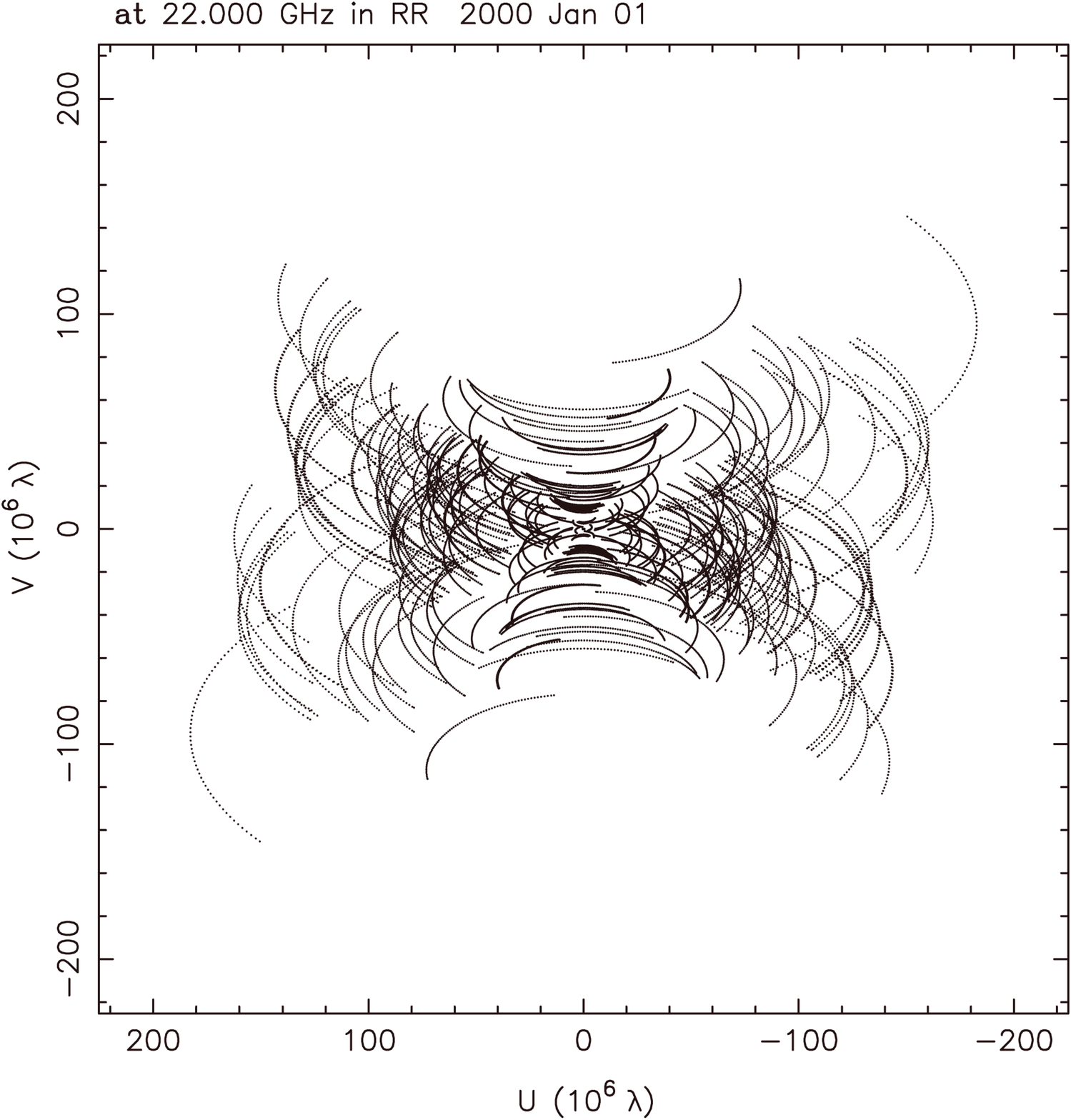}
{fig:Figure2}
{Examples of ($u$, $v$) coverage for an EAVN observation at 22~GHz with the
source's declination of $+60^{\circ}$ (top left), $+30^{\circ}$ (top right),
$0^{\circ}$ (bottom left), and $-29^{\circ}$ (bottom right). Total observation
duration of 10~hours is assumed for all cases.}

\section{Results of Fringe Test Observations}

We have carried out test observations with the EAVN eight times since 2013
at 8 and 22~GHz in order to get fringes between international baselines of
the EAVN, and to clarify the problems associating with the scheduling,
correlation, and data reduction.
16 telescopes have participated in fringe tests one or more times at either
8 or 22~GHz, or at both frequencies (see Table~\ref{tbl:Table1}).

\articlefiguretwo
{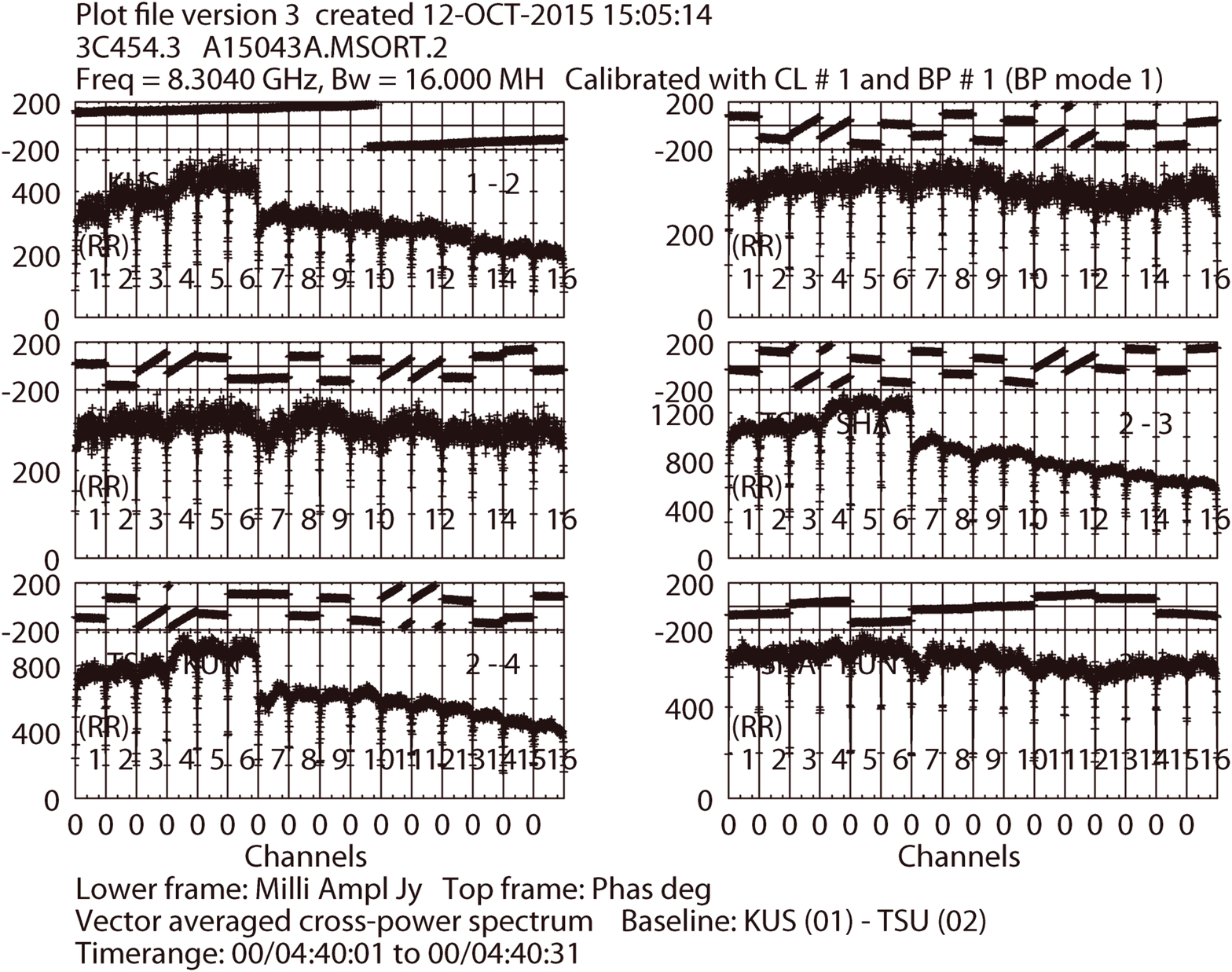}
{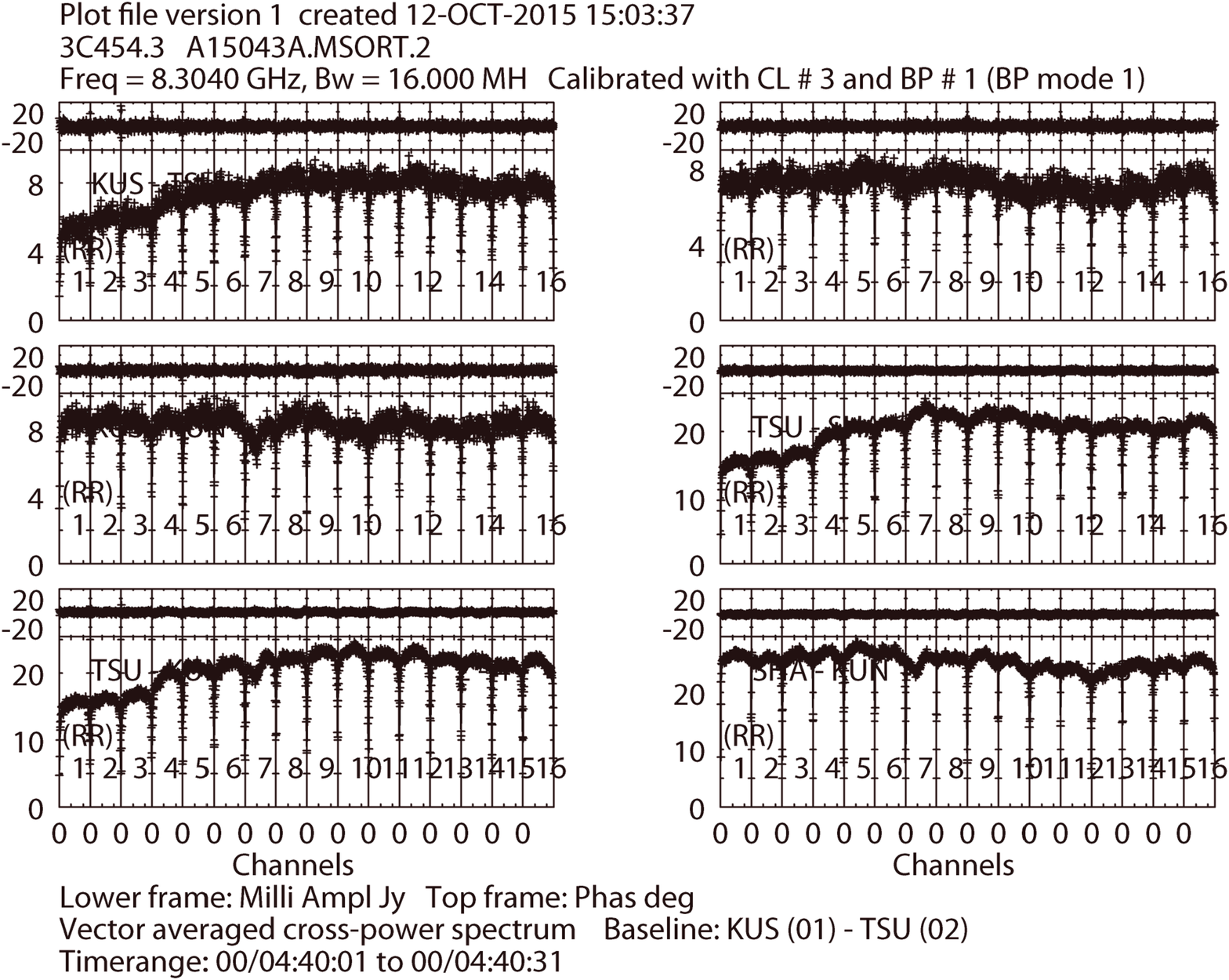}
{fig:Figure3}
{Results of the fringe test observation conducted on 2015 February 12 at
8~GHz. (Left) Raw visibility phases (upper) and amplitudes (lower) for each
baseline. (Right) Calibrated visibility phases and amplitudes with the AIPS
task FRING and ACCOR. Note that {\it a~priori} amplitude calibration with
the AIPS task APCAL is not applied.}

Throughout those fringe tests, we could detect fringes at both frequencies.
Figure~\ref{fig:Figure3} shows examples of fringe test results conducted
on 2015 February 12 at 8~GHz.
Four telescopes, Tsukuba (JVN), Ulsan (KVN), Sheshan and Kunming (CVN),
participated in the observation, with a bright AGN 3C~454.3 being a target
source.
Fringes were successfully detected between all international baselines in
China, Japan, and Korea.
Although discontinuous fringe phases and amplitudes can be seen between
each IF channel in the raw visibility, those are well-calibrated after
fringe fitting and amplitude correction.
Discontinuous fringe phases appear in the international baselines between
trans-CVN baselines (i.e., Sheshan/Kunming -- Tsukuba/Ulsan baselines),
although those cannot be seen in the visibility of Sheshan -- Kunming and
Tsukuba -- Ulsan baselines.
This phenomenon is probably due to the different bit assignment in the
backend system between CVN and KVN/JVN stations.
This can be compensated by the correlator and good fringe phases and amplitudes
can be obtained after fringe fitting and amplitude calibration, as shown
in Figure~\ref{fig:Figure3}.

\section{Future Plan and Synergy with the FAST}

\subsection{Future Plan of the EAVN Test Observation}

On the basis of the results obtained with eight-time fringe tests, we
will start imaging test observations with the EAVN from the end of 2015.
We conduct test observations at 8 and 22~GHz as a first step, whereas
we will also begin both fringe tests and imaging observations from
2016 at 6.7 and 43~GHz, both of which are common observing frequencies
and science commissioning observations have already been conducted with
part of EAVN telescopes, as mentioned in Section~1.
As of this stage, we will begin to invite observing proposals for
the EAVN from the second half of 2017.

\subsection{Synergy and Collaboration with the FAST/Other Arrays}

The FAST telescope will have a capability of the data reception system
at 70~MHz -- 3~GHz, while some of EAVN telescopes have receivers at
1.6 and/or 2~GHz.
The minimum sensitivity for 7-$\sigma$ fringe detection of a few hundred
mJy will be achieved at both 1.6 and 2~GHz between the FAST and the Usuda
64~m telescope of the JVN, assuming the antenna diameter of 300~m, the
aperture efficiency of 20\%, and the system noise temperature of 25~K for
the FAST, and the integration time of 60 seconds.
The extremely high sensitivity allows us to investigate relatively weak
radio sources such as low-luminosity AGNs, and to detect much more
OH (mega-)maser sources in galactic and extragalactic objects.

In the Asia-Pacific region, the Long Baseline Array (LBA) in Australia
has already been made regular operation, and collaborative work between
the EAVN and the LBA has been made as a part of the framework in the
Asia-Pacific Telescope (APT).
The maximum baseline length of longer than 10,000~km can be obtained with
the combined EAVN and APT array in the north-south direction.
This enable us to obtain better angular resolution and ($u$, $v$) coverage
toward sources at low declination and southern sources.
A new VLBI array project, the Thai VLBI Network (TVN), is also in
progress in Thailand, and we are planning to pursue collaboration with
both LBA and TVN in the near future,


\end{document}